\documentclass[aps,prb,superscriptaddress, twocolumn]{revtex4-1}
\usepackage{amsmath}
\usepackage{epsfig}
\usepackage{graphicx}
\usepackage{color}
\usepackage{multirow}
\usepackage{array}
\usepackage{tabularx}
\usepackage{inputenc}
\newcommand{\rt}{R$_2$T$_2$O$_7$}
\newcommand{\erti}{Er$_2$Ti$_2$O$_7$}
\newcommand{\er}{Er$^{3+}$}

\newcolumntype{M}[1]{>{\centering\arraybackslash}m{#1}}
\newcolumntype{Y}{>{\centering\arraybackslash}X}

\begin{document}
\title{Field induced phase diagram of the XY pyrochlore antiferromagnet \erti}

\author{E. Lhotel}
\email[]{elsa.lhotel@neel.cnrs.fr}
\affiliation{Institut N\'eel, CNRS and Universit\'e Grenoble Alpes, 38042 Grenoble, France}
\author{J. Robert}
\affiliation{Institut N\'eel, CNRS and Universit\'e Grenoble Alpes, 38042 Grenoble, France}
\author{E. Ressouche}
\affiliation{INAC, CEA and Universit\'e Grenoble Alpes, CEA Grenoble, 38054 Grenoble, France}
\author{F. Damay}
\affiliation{Laboratoire L\'eon Brillouin, CEA, CNRS, Universit\'e Paris-Saclay, CE-Saclay, 91191 Gif-sur-Yvette, France}
\author{I. Mirebeau}
\affiliation{Laboratoire L\'eon Brillouin, CEA, CNRS, Universit\'e Paris-Saclay, CE-Saclay, 91191 Gif-sur-Yvette, France}
\author{J. Ollivier} 
\affiliation{Institut Laue Langevin, CS 20156, 38042 Grenoble, France}
\author{H. Mutka} 
\affiliation{Institut Laue Langevin, CS 20156, 38042 Grenoble, France}
\author{P. Dalmas de R\'eotier}
\affiliation{INAC, CEA and Universit\'e Grenoble Alpes, CEA Grenoble, 38054 Grenoble, France}
\author{A. Yaouanc}
\affiliation{INAC, CEA and Universit\'e Grenoble Alpes, CEA Grenoble, 38054 Grenoble, France}
\author{C. Marin}
\affiliation{INAC, CEA and Universit\'e Grenoble Alpes, CEA Grenoble, 38054 Grenoble, France}
\author{C. Decorse}
\affiliation{ICMMO, Universit\'e Paris-Saclay, Universit\'e Paris-Sud, 91405 Orsay, France}
\author{S. Petit}
\email[]{sylvain.petit@cea.fr}
\affiliation{Laboratoire L\'eon Brillouin, CEA, CNRS, Universit\'e Paris-Saclay, CE-Saclay, 91191 Gif-sur-Yvette, France}
\begin{abstract}
We explore the field-temperature phase diagram of the XY pyrochlore antiferromagnet \erti\, by means of magnetization and neutron diffraction experiments. Depending on the field strength and direction relative to the high symmetry cubic directions $[001], [1\bar{1}0]$ and $[111]$, the refined field induced magnetic structures are derived from the zero field $\psi_2$ and $\psi_3$ states of the $\Gamma_5$ irreducible representation which describes the ground state of XY pyrochlore antiferromagnets. At low field, domain selection effects are systematically at play. In addition, for $[001]$, a phase transition is reported towards a $\psi_3$ structure at a characteristic field $H_c^{001}=$ 43 mT. For $[1\bar{1}0]$ and $[111]$, the spins are continuously tilted by the field from the $\psi_2$ state, and no phase transition is found while domain selection gives rise to sharp anomalies in the field dependence of the Bragg peaks intensity. For $[1\bar{1}0]$, these results are confirmed by high resolution inelastic neutron scattering experiments, which in addition allow us to determine the field dependence of the spin gap. This study agrees qualitatively with the scenario proposed theoretically by Maryasin {\it et al.} [Phys. Rev. B {\bf 93}, 100406(R) (2016)], yet the strength of the field induced anisotropies is significantly different from theory.

\end{abstract}
%
\maketitle

\section{Introduction}
The family of pyrochlore compounds \rt\, (R is a rare earth and T=Ti, Sn, Zr, ...) has aroused a lot of interest in the last years \cite{gingrasrmp}. In these materials, the R magnetic moments reside on the vertices of corner sharing tetrahedra, a configuration especially relevant to study magnetic geometric frustration \cite{Lacroix,bramgin00}. In this context, the case of \erti\, is noteworthy as it may be one of the scarce examples where the frustration is resolved by an ``order by disorder'' phenomenon \cite{Champion03,Savary12,Zhitomirsky12}. This is a degeneracy lifting mechanism where thermal or quantum fluctuations select the ground state from a classically degenerate manifold that has the largest space to fluctuate \cite{Villain,Shender}. Recently, this picture has been questioned: the relevance of an energetic selection mechanism that would proceed through ``virtual crystal field'' excitations of the \er\, ion has been pointed out (VCF model) \cite{McClarty09,Petit14,Rau16}. It remains however difficult to distinguish between the different mechanisms. 
Depending on the direction along which it is applied, a magnetic field will compete, or not, with this zero field selection mechanism. This results in a rich field-temperature phase diagram, especially at low field \cite{Maryasin16}, which is intimately related to the system's microscopic parameters. In this article, this physics is explored experimentally in a systematic way.

\begin{figure}[!t]
\includegraphics[width=9cm]{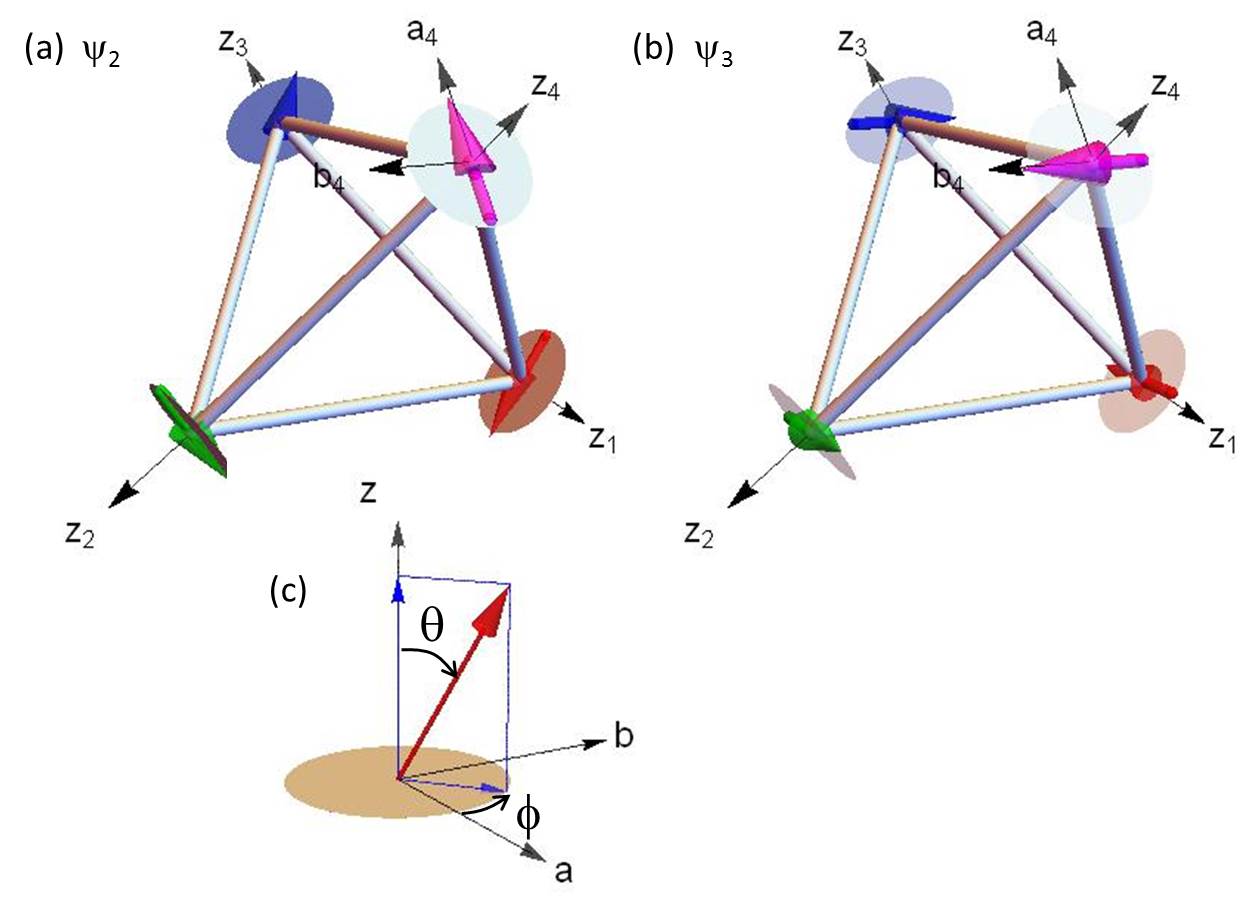}
\caption{(Color online) (a,b) Sketch of one tetrahedron of the pyrochlore structure in the $\psi_2$ (a) and $ \psi_3$ (b) magnetic configurations. $\psi_2$ and $\psi_3$ are the two basis vectors of the $\Gamma_5$ irreducible representation. 
The disks represent the local XY anisotropy planes. The black arrows denote the CEF ${\boldsymbol z_i}$ axes. For the site $i=4$, the ${\boldsymbol a_i}$ and ${\boldsymbol b_i}$ axes are also shown.
(c) Sketch defining the $\phi$ angle within the XY $({\boldsymbol a},{\boldsymbol b})$ plane and the polar angle $\theta$ relative to the CEF axis ${\boldsymbol z}$.}
\label{Figure1}
\end{figure}

\erti\, was identified as an XY antiferromagnet \cite{Champion03}, the spin being confined by the crystal electric field (CEF) within a site dependent local XY anisotropy plane, defined by $({\boldsymbol a_i},{\boldsymbol b_i},{\boldsymbol z_i})$ vectors, with ${\boldsymbol z_i}$ the local CEF axis (see Table \ref{table-sym} for the appropriate definition). Among the possible states that minimize the XY anisotropy energy, the $\Gamma_5$ irreducible representation manifold (see Figure \ref{Figure1}) is expected to be the ground state for isotropic antiferromagnetic exchange. It possesses a U(1) degeneracy, any configuration of the form ${\boldsymbol S_i} = \cos \phi~{\boldsymbol a_i} + \sin \phi ~{\boldsymbol b_i}$ being a possible classical ground state \cite{Champion03,Savary12,Zhitomirsky12}. Note that $\phi$ is an arbitrary angle but is the same for all spins. 

On the experimental side, \erti\, orders below $T_{\rm N}=1.2$~K \cite{Blote69,Harris98, Siddharthan99,Poole07} in the non-collinear so called $\psi_2$ magnetic structure, depicted in Figure \ref{Figure1}(a), and characterized by the 6 domains defined by $\phi=n \pi/3, n=0,..,5$ (see Table \ref{table-sym}). 

\begin{table}
\setlength{\extrarowheight}{1pt}
\begin{tabularx}{8.5cm}{M{1.75cm}M{1.75cm}YM{1.5cm}Y} 
\hline
\hline
Site & 1 & 2 & 3 & 4 \\
CEF axis ${\boldsymbol z_i}$ & $(1,1,\bar{1})$ & $(\bar{1},\bar{1},\bar{1})$ & $(\bar{1},1,1)$ & $(1,\bar{1},1)$ \\
Coordinates & {\text{\footnotesize$(1/4,1/4,1/2)$}} & {\text{\footnotesize$(0,0,1/2)$}} & {\text{\footnotesize$(0,1/4,3/4)$}} & {\text{\footnotesize$(1/4,0,3/4)$}}\\
\hline 
${\boldsymbol a_i}$ & $(\bar{1},\bar{1},\bar{2})$ & $(1, 1,\bar{2})$ & $(1,\bar{1},2)$ & $(\bar{1},1,2)$ \\
${\boldsymbol b_i}$ & $(\bar{1},1,0)$ & $(1, \bar{1},0)$ & $(1,1,0)$ & $(\bar{1},\bar{1},0)$ \\
\hline
$n=0$ & $(\bar{1},\bar{1},\bar{2})$ & $(1, 1,\bar{2})$ & $(1,\bar{1},2)$ & $(\bar{1},1,2)$ \\
$n=2$ & $(\bar{1},2,1)$ & $(1,\bar{2},1)$ & $(1,2,\bar{1})$ & $(\bar{1},\bar{2},\bar{1})$ \\
$n=4$ & $(2,\bar{1},1)$ & $(\bar{2},1,1)$ & $(\bar{2},\bar{1},\bar{1})$ & $(2,1,\bar{1})$ \\
\hline
\hline
\end{tabularx}
\caption{Coordinates, written in the cubic $Fd\bar{3}m$ structure of the pyrochlore lattice, of the site dependent local ${\boldsymbol a_i}$ and ${\boldsymbol b_i}$ vectors spanning the local XY anisotropy planes. The CEF axes ${\boldsymbol z_i}$ of the rare earth are perpendicular to those planes. The coordinates of the spin direction in each of the 6 $\psi_2$ domains for $\phi=0, 2\pi/3$ and $4\pi/3$ ($n=0, 2$ and $4$) are also given. The domains corresponding to $\phi=\pi, 5\pi/3, \pi/3$ ($n=3, 5$ and $1$) are obtained by taking the opposite vectors. }
\label{table-sym}
\end{table}

It has been recognized quite soon that the reason for the stabilization of $\psi_2$, hence for the U(1) degeneracy breaking, could be due to thermal and quantum order by disorder \cite{Champion03,Savary12,Zhitomirsky12,Yan13,Wong13,Oitmaa13,McClarty14,Zhitomirsky14,Maryasin14,Javanparast15}. 
In that context, Savary {\it et al.} \cite{Savary12} have described \erti\ by a generic quadratic Hamiltonian ${\cal H}$ written in terms of the components of an effective spin $1/2$ spanning the subspace of the ground \er\, crystal field doublet \cite{Curnoe08,Ross11,Onoda11}:
\begin{eqnarray*}
{\cal H} &=& \sum_i g~{\sf S}_i.{\boldsymbol H} + \frac{1}{2} \sum_{i,j} {\sf J}_{zz} {\sf S}^z_i {\sf S}^z_j + {\sf J}_{z \pm}{\sf S}_i^z \left( \zeta_{ij} {\sf S}^+_j + \zeta^*_{ij} {\sf S}^-_j\right) \\
& &
+ 
{\sf J}_{\pm\pm} \left(\gamma_{ij} {\sf S}^+_i {\sf S}^+_j + \gamma^*_{ij} {\sf S}^-_i {\sf S}^-_j \right) - {\sf J}_{\pm} \left({\sf S}^+_i {\sf S}^-_j + {\sf S}^-_i {\sf S}^+_j \right)
\end{eqnarray*}
${\sf S}_i$ denotes the pseudo spin $1/2$ written in its local basis $({\boldsymbol a_i},{\boldsymbol b_i},{\boldsymbol z_i})$, $g$ is an effective anisotropic tensor, $\gamma_{ij}$ and $\zeta_{ij}$ are complex unimodular matrices, and $({\sf J}_{\pm\pm},{\sf J}_{\pm},{\sf J}_{z\pm},{\sf J}_{zz})$ is a set of effective exchange parameters allowed by symmetry and determined based upon fitting the spin wave excitations \cite{Savary12} (see Appendix \ref{models}). Using these parameters, Savary {\it et al.} could predict a gap in the spin excitation spectrum of about 20 $\mu$eV \cite{Savary12}, comparable to the 43 $\mu$eV gap that was observed experimentally by inelastic neutron scattering \cite{Ross14,Petit14} and the upper bound of which was determined from electron paramagnetic resonance \cite{Sosin10} and specific heat \cite{Reotier12} measurements. 
 
Concurrently, it was shown that an energetic selection mechanism that proceeds through virtual crystal field excitations can be at play, giving rise to a gap of about 10 $\mu$eV \cite{McClarty09,Petit14,Rau16}. 
Moreover, it was pointed out that once projected onto the effective spin $1/2$ subspace, the virtual crystal field excitations of the \er\, ion give rise to multispin interactions like for instance a biquadratic coupling \cite{Rau16}. It has been proposed that the discrepancy between the theoretical and experimental gap could be resolved with reasonable values of this coupling, maintaining an excellent agreement with the spin excitation spectrum \cite{Rau16}. This approach appears particularly relevant since the importance of such multispin interactions has been recently put forward for other pyrochlores \cite{ndzr, przr, tbti}.

Maryasin {\it et al.} \cite{Maryasin16} have shown that applying a magnetic field ${\boldsymbol H}$ along the high symmetry directions of the cubic lattice reveals a rich field-temperature $H-T$ phase diagram, characterized by the competition between the zero-field selection mechanism and field induced anisotropies. The strength of these anisotropies is directly related to the microscopic parameters of ${\cal H}$, so that the experimental determination of the phase diagram should allow to go a step further in the understanding of \erti. 

In this paper, we address this phase diagram through magnetization and neutron scattering measurements and determine the field induced magnetic structures. The obtained characteristic fields can then be compared with the theoretical predictions of Ref. \onlinecite{Maryasin16}.

The paper is organized as follows: we first discuss briefly the field evolution of the spin dynamics in the ``high field" regime, i.e. at fields large enough to overcome the anisotropy terms, so that the system is nearly polarized by the magnetic field. We then turn to the evolution of the magnetic structure when a magnetic field is applied along the three high symmetry directions, focusing on the ``low field" part of the phase diagram.
We introduce the theoretical predictions and then describe our experimental results. Finally, the similarities and differences between theory and experience are discussed. 

\begin{figure}[!]
\includegraphics[height=8cm]{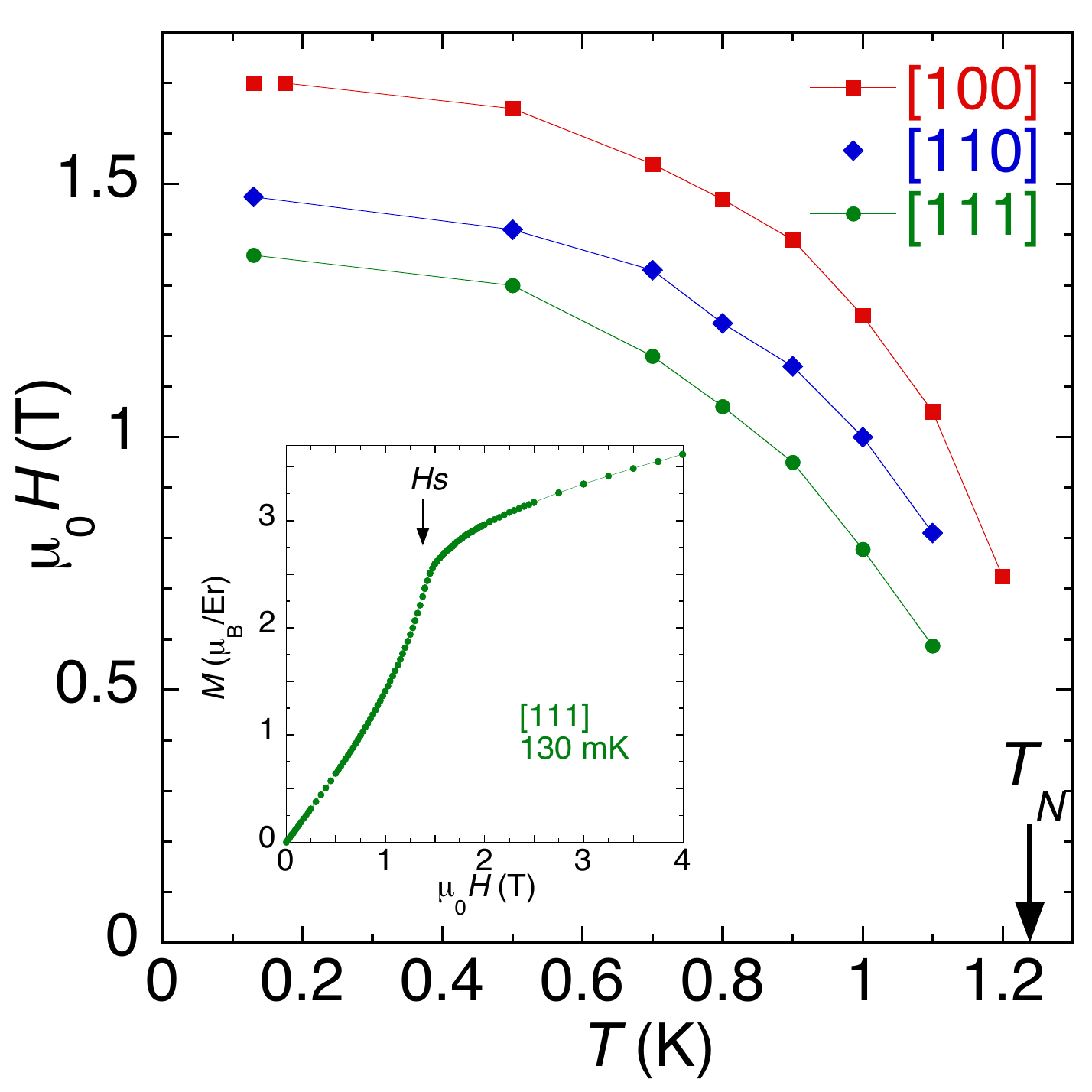}
\caption{(Color online) Field - temperature $(H, T)$ phase diagram obtained for a field applied along the three main directions of the cubic lattice. Inset: $M$ vs $H$ at 130 mK, and ${\boldsymbol H} \parallel [111]$ showing the critical field at the inflection point of the magnetization curve.}
\label{pdiag}
\end{figure}

\section{Experimental details}
\begin{figure*}[!ht]
\includegraphics[height=6.5cm]{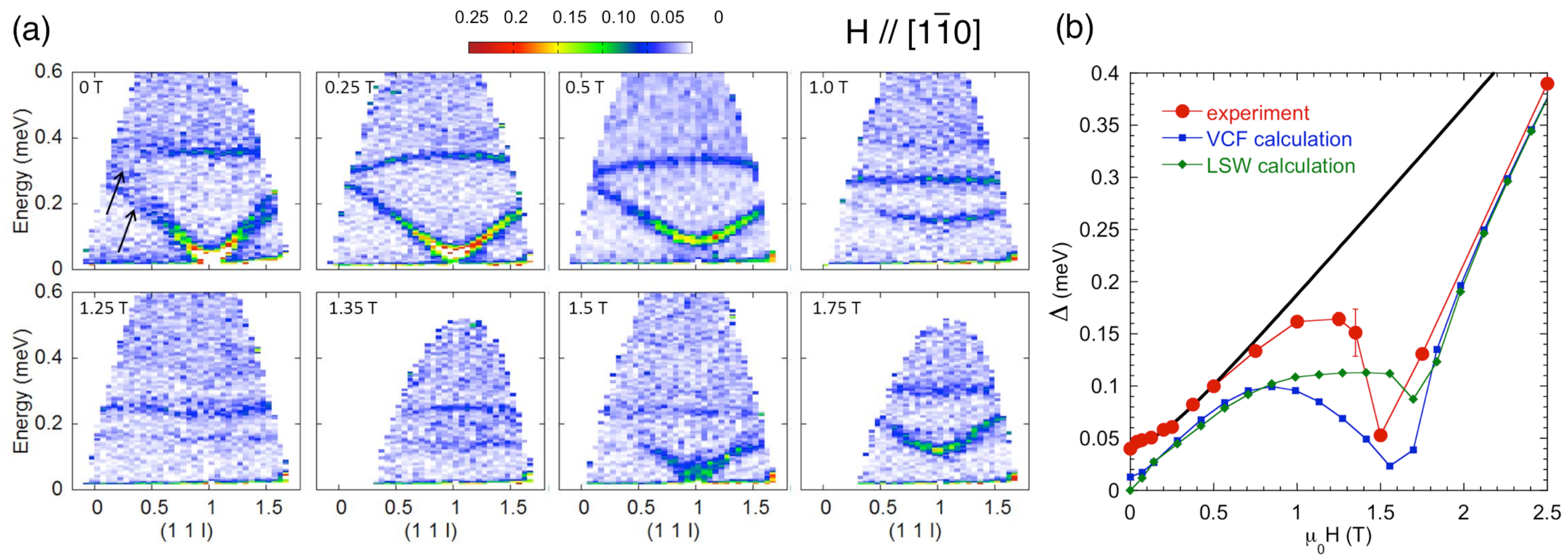}
\caption{(Color online) (a) Inelastic neutron scattering data recorded along $(11\ell)$ as a function of magnetic field applied along $[1\bar{1}0]$ at 60 mK on IN5. The spin gap is determined at the magnetic zone center $Q=(111)$. Arrows in the zero field data point the acoustic branches arising from the different domains. (b) Spin gap $\Delta$ as a function of $H$. Red points correspond to the values determined from a Lorentzian fit to the data. Blue squares and green diamonds correspond respectively to the gap obtained from RPA calculations in the virtual crystal field (VCF) model \cite{Petit14} and from the linear spin wave (LSW) theory using ${\cal H}$ \cite{Savary12}. The black line is a fit to the equation $\sqrt{a+bH^2}$ with $a=1.96 \pm 0.02 \times 10^{-3}$ meV$^{2}$ and $b=0.033 \pm 0.003$ meV$^{2}/$T$^{2}$, for $\mu_0H<0.5$ T.}
\label{Figure-INS}
\end{figure*}

Experiments were performed on single crystals of \erti\ grown by the floating zone technique. The samples used for neutron scattering measurements come from a large \erti\, single crystal previously used in Ref. \onlinecite{Petit14} and \onlinecite{Cao10} and cut in different pieces depending on the type of experiment. Magnetization measurements were performed on a small piece of the crystal of Ref. \onlinecite{Reotier12}. 

Inelastic neutron scattering experiments were carried out on the IN5 disk chopper time of flight spectrometer operated by the Institute Laue Langevin (ILL France). The field was applied along $[1\bar{1}0]$ and the sample mounted to have the $(hh0)$ and $(00\ell)$ reciprocal directions in the horizontal scattering plane. As a very good energy resolution, about 20 $\mu$eV, is needed to observe the spin gap, we used a wavelength $\lambda=8.5$ \AA. The data were then processed with the {\sc horace} software \cite{horace}, transforming the recorded time of flight, sample rotation and scattering angle into energy transfer and $Q$-wave-vectors. The offset of the sample rotation was determined based on the Bragg peaks positions. In all the experiments, the sample was rotated by steps of 1 degree.

The neutron diffraction data were collected using the D23 single crystal diffractometer (CEA-CRG, ILL France) operated with a copper monochromator and using $\lambda=1.28$ \AA. The \erti\, sample was glued on the Cu finger of a dilution insert and placed in a cryomagnet. The experiments have been conducted with the (vertical) field ${\boldsymbol H}$ either parallel to the $[001]$, $[1\bar{1}0]$ or $[111]$ high symmetry crystallographic directions. 
Since the magnetic structure is ${\boldsymbol K} = {\bf 0}$, additional magnetic intensity is expected on crystalline Bragg peaks. For reference, a series of integrated intensities at 10 K and $H=0$ was measured (about 70 reflections). Refinements were performed using the {\sc Fullprof} sofware suite \cite{fullprof}, working on the subtraction with the $T=10$~K data. In addition, for a set of chosen Bragg peaks, the intensity versus sweeping the magnetic field (0.015~T/min was the minimum speed) was collected, to obtain the precise evolution of the magnetic intensity. For all the measurements, specific care was taken to apply systematically the same field sweeping conditions, to avoid possible irreversibilities. The sample was first cooled down to 60 mK in zero field. All the measurements were then performed after the application of a 3 T field used to saturate the sample. The shape of the samples was not optimized regarding the demagnetizing effects, leading to relatively large demagnetizing factors, estimated between 5 and 10 (cgs units), depending on the orientation with respect to the magnetic field. The applied magnetic field was then significantly larger than the internal field. We have not performed the demagnetization corrections which would have led to sizable uncertainties.

Magnetization and ac susceptibility measurements were performed down to 100~mK on a single crystal sample using a superconducting quantum interference device (SQUID) magnetometer equipped with a dilution refrigerator developed at the Institut N\'eel-CNRS Grenoble \cite{Paulsen01}. The sample had a flat ellipsoid shape and the field was applied in the disk plane so as to minimize the demagnetization effects. Magnetization curves were measured using a 0.5 mT field step. 

\section{Spin dynamics below the field polarized state}
\label{HF} 
The high field phase diagram of \erti\, is well documented and has been characterized by neutron scattering, magnetization and specific heat measurements \cite{Ruff08,Cao10,Petrenko13,Bonville13,Reotier12,Bertin12}. Figure \ref{pdiag} shows this phase diagram, obtained from magnetization measurements. At high field, above a critical field $H_S$, the system is in a field polarized state. This critical field depends on the orientation of the field: 1.7~T for ${\boldsymbol H} \parallel [001]$, 1.5 T for ${\boldsymbol H} \parallel [1\bar{1}0]$ and 1.35 T for ${\boldsymbol H} \parallel [111]$ at 100 mK. 

The field also induces a change of the spin dynamics, as shown in Figure \ref{Figure-INS} for ${\boldsymbol H} \parallel [1\bar{1}0]$. In this case, the applied field selects the $\psi_2$ domains with $\phi=0$ and $\pi$ (among the 6 present in zero field). A signature of this domain selection in the excitation spectrum is the vanishing of the acoustic branches arising from the unfavored domains. This is visible when comparing the $\mu_0 H=0.25$~T with the zero field data for instance (see the arrows on Figure~\ref{Figure-INS}(a) at 0 T). 

When the field is further increased, the spins are gradually tilted, before reaching the critical field $H_S$. The latter manifests as a cusp in the intensity of certain Bragg peaks, as shown in the diffraction measurements and the field induced magnetic structures in Section \ref{LF}. 
Our inelastic measurements evidence that below $H_S$ the spectrum resembles the zero field spectrum, with regular spin waves, except that the spin gap increases, and that some intensity is lost when approaching $H_S$. These results clarify the previous description of Ruff {\it et al.} \cite{Ruff08} (see also Ref. \onlinecite{Gaudet16}) who could not detect the dispersion of the high energy mode. At $H_S$, the gap vanishes and the dispersion of the field polarized phase is recovered. 

An interesting issue is the field dependence of the spin gap $\Delta$. Fitting the spin wave modes at $Q=(111)$ to Lorentzian profiles, we obtain $\Delta$ as a function of $H$ (see Figure \ref{Figure-INS} (b)). Starting from $\Delta_o \approx 43~\mu$eV, it reaches $160~\mu$eV at about 1.2 T and then decreases down to zero at the saturation field $\mu_0H_S=$ 1.5 T. 

\begin{figure}[t!]
\includegraphics[width=8cm]{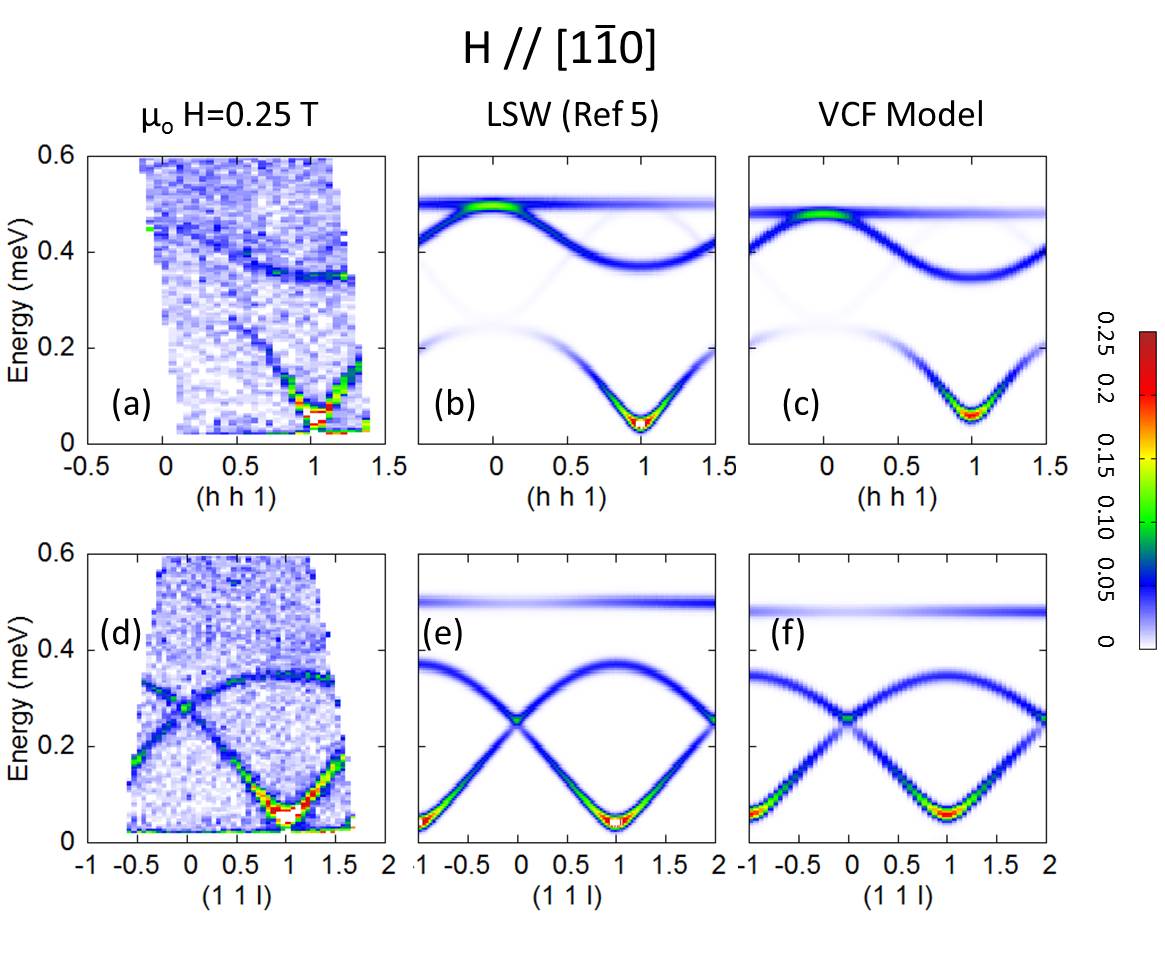}
\caption{(Color online) Measured and calculated inelastic neutron scattering spectra at $\mu_0H=0.25$ T along $[1\bar{1}0]$ for the $(hh1)$ and $(11\ell)$ directions. The calculations are performed for the two approaches (LSW and VCF) described in the text.}
\label{Figure-H0p25T}
\end{figure}

The measured field dependence of the spin gap has been compared with two models. The first one, the VCF model, is based on a mean field treatment of the full Hamiltonian written for the actual \er\, ions and is described in Ref. \onlinecite{Petit14}. It takes into account explicitly the CEF Hamiltonian \cite{wybourne} (with the parameters of Cao {\it et al.} \cite{Cao09,bnm}, see also Ref. \onlinecite{Bonville13, Cao10, Bertin12}) and calculates the spin dynamics in the  Random Phase Approximation (RPA) \cite{jensen}. The second one is a linear spin wave (LSW) calculation \cite{spinwave} based on the spin 1/2 Hamiltonian ${\cal H}$ and on the parameters determined in Ref. \onlinecite{Savary12} (see Appendix~\ref{models}). These two approaches reproduce successfully the experimental spectra as shown in Figure \ref{Figure-H0p25T}. Nevertheless, in zero field, the spin gap is underestimated in the VCF model \cite{Rau16} and (as expected) is zero in the LSW theory. In addition, while the qualitative behavior of the field dependence of the spin gap is well captured by both models, none is able to reproduce the strong increase of $\Delta$ at intermediate fields, just before the saturation at $H_S$ (see Figure~\ref{Figure-INS}). This discrepancy suggests that some ingredients are missing in the Hamiltonian to describe accurately the field behavior. 
 Accounting for additional interactions, such as multispin ones, which have been proposed in the context of the VCF model \cite{Rau16}, may improve the agreement between models and experiments.


\section{Field evolution of the magnetic structures}
 \label{LF}
In this section, we turn to the description of the phase diagram by focusing mainly on the ``low field" part, i.e. when the field is small enough not to tilt the magnetic moments with respect to their easy anisotropy plane. This region of the phase diagram has been little explored experimentally.
Anomalies have been reported in the isothermal magnetization curves below 0.1~T \cite{Petrenko13}, which appear as maxima in the derivative of the magnetization with respect to the magnetic field $dM/ dH$, and indicate that field induced transitions may occur at low field. 
Subsequently, Maryasin et al. \cite{Maryasin16} have performed a detailed analysis of this phase diagram by considering the anisotropy terms authorized by symmetry in presence of a magnetic field. These new terms can possibly compete with the zero field term responsible for the $\psi_2$ magnetic ordering. Maryasin et al. have made quantitative theoretical predictions which, if verified, would allow one to determine the value of these anisotropy parameters. 

We first summarize the main theoretical results obtained in Ref. \onlinecite{Maryasin16} when the field is applied along the three directions of high symmetry $[001]$, $[1\bar{1}0]$ and $[111]$, and then describe our experimental results.\\

\begin{table*}[t!]
\setlength{\extrarowheight}{1pt}
\begin{tabularx}{\textwidth}{M{3cm} M{4.8cm} *{2}{Y}}
\hline \hline
${\boldsymbol H}$ direction&Prediction& Predicted characteristic field & Observed characteristic field \\  \hline
$[001]$ & Transition $\psi_2 \rightarrow \psi_3$& $H_c^{[001]}=3 \sqrt{A_6/A_2}$ & 43 mT \\
$[1\bar{1}0]$ & Domain selection within $\psi_2$ & $H^{[1\bar{1}0]}=3 \sqrt{2A_6/A_2}$ & 74 mT \\
\multirow{2}{*}{$[111]$} & Domain selection within $\psi_2$ only & &100 mT \\
&or intermediate phase & & not observed \\ \hline \hline
\end{tabularx}
\caption{\label{summary} Summary of the field induced behaviors expected in the low field regime, from the calculations of Ref. \onlinecite{Maryasin16}.} 
\end{table*}

\subsection{Theoretical background}
\label{theory}

Whatever its actual physical origin, the lifting of the ground state degeneracy, which is responsible for the stabilization of an ordered $\psi_2$ state, can indeed be accounted for by adding an effective 6-fold anisotropy term to the free energy \cite{Zhitomirsky14}: $$\Delta F_0=-A_6 \cos 6 \phi$$ where $\phi$ is the angle defined in the Introduction part and which defines the magnetic structures within the $\Gamma_5$ manifold. $A_6$ is a parameter depending on the microscopic details of the Hamiltonian ${\cal H}$. $A_6 > 0$ ensures that $\psi_2$ is the ground state, with minima at $\phi=n\pi/3$, thus stabilizing the 6 magnetic domains described above. 

In the presence of a magnetic field ${\boldsymbol H}$, new terms arise in the free energy. Following a symmetry analysis, Maryasin {\it et al.} have established the general form of the lowest order terms \cite{Maryasin16}. New parameters (called $A_i$ or $A'_i$) are introduced which are related to the Hamiltonian and are all positive in the case of \erti. The first terms which arise are quadratic in $H$: 
\begin{align}
\label{df2}
\Delta F_2 = &A_6' H^2 \cos 6\phi \quad \\
+ &A_2 {\begin{aligned}[t] \left[ \left(H_z^2-\frac{H_x^2}{2}-\frac{H_y^2}{2}\right) \cos 2\phi \right. \\
\left. - \frac{\sqrt{3}}{2}(H_x^2-H_y^2) \sin 2\phi \right] \end{aligned}} \notag 
\end{align}
where the first $A'_6$ term is usually small compared to the second $A_2$ one. 

The next term, which is generally negligible, is at third order in $H$ and writes: 
\begin{equation}
\Delta F_3 =3 \sqrt{3} A_3 H_x H_y H_z \cos 3\phi
\label{df3}
\end{equation} 

When the field is smaller than the saturating field $H_S$ (see Section \ref{HF}), the field behavior will be extremely different depending on the values of the coefficients $A_i$. In particular, these coefficients determine which of the $\psi_2$ domains are favored by the field, as well as a complex phase diagram that depends on the field orientation. A careful study of the low field behavior will then allow to access directly to these coefficients. Based on Ref. \onlinecite{Maryasin16}, we detail below the main characteristic of the field induced properties that result from the above free energy when $H<H_S$ (see also Table \ref{summary} for a summary). 

\subsubsection{${\boldsymbol H} \parallel [001]$} 
For ${\boldsymbol H} \parallel [001]$, and considering only the $A_2$ term in equation (\ref{df2}), the free energy writes: 
$$\Delta F = - A_6 \cos 6\phi + A_2 H^2 \cos 2\phi $$
The two terms of $\Delta F$ compete with each other when the field increases. For magnetic fields up to $H < H^{001}_c$, with
\begin{equation} 
H^{001}_c=3 \sqrt{A_6/A_2}
\label{hc001}
\end{equation}
the energy is minimized for four domains out of six with $\phi$ varying continuously when the field increases, and which originate from the $n=1,2,4,5$ domains of the $\psi_2$ state. 

Above this critical value, a transition is expected towards a $\psi_3$ state, with two domains corresponding to $\phi= \pm\pi/2$, until the system reaches the polarized state described in the previous section at $H_S^{001}$. 

\subsubsection{${\boldsymbol H} \parallel [1\bar{1}0]$}

For ${\boldsymbol H} \parallel [1\bar{1}0]$, and considering the same terms in the free energy,
$$\Delta F = - A_6 \cos 6\phi - \frac{1}{2} A_2 H^2 \cos 2\phi $$ so that both terms favor the $\psi_2$ state until the field polarized state is achieved at $H^{110}_S$. Nevertheless a domain selection is at play and the two domains with $\phi=0,\pi$ (i.e. $n=0,3$, as discussed in section \ref{HF}) are selected when the field reaches the value $H^{110}$, with
\begin{equation} 
H^{110}=3 \sqrt{2A_6/A_2}= \sqrt{2} {H^{001}_c}
\label{hc110}
\end{equation}

\subsubsection{${\boldsymbol H} \parallel [111]$} 
For ${\boldsymbol H} \parallel [111]$, the $A_2$ contribution in $\Delta F_2$ vanishes and the third order $A_3$ term (equation (\ref{df3})) becomes relevant, along with the $A'_6$ term of equation (\ref{df2}). The free energy then writes: 
$$\Delta F =-A_6 \cos 6 \phi+ A_6' H^2 \cos 6\phi + A_3 H^3 \cos 3\phi$$
In that case, the discussion is more complex and depends on the ratio $\zeta= \dfrac{4(A_6'H^2-A_6)}{A_3H^3}$. 

For $\zeta < 1$, three domains of the $\psi_2$ state, corresponding to $\phi=n\pi/3$ with $n=1,3,5$, are favored by the field, until $H_S^{111}$ is reached. 
 
Interestingly, for $\zeta > 1$, while the same selection occurs at low field and just below $H_S^{111}$, a new phase is predicted at intermediate fields. This intermediate phase is based on the $\psi_3$ states and arises from the competition between $A'_6$ and $A_6$. It is essentially stabilized in the field region where $A'_6 H^2$ overcomes $A_6$. 

The microscopic parameters estimated for \erti\, place the system close to the boundary between these two regimes, with $\zeta$ slightly larger than one, so that the existence of this intermediate phase is an open question at the moment. Note however that when the temperature increases, the six-fold anisotropy $A_6$ of $\Delta F_0$ is reinforced by thermal fluctuations, so that the $\psi_2$ state is expected to be stable in the whole field range below $H_S^{111}$.

\subsection{Experimental results}
To determine the changes in the magnetic structures corresponding to the anomalies reported at low field in the isothermal magnetization curves \cite{Petrenko13}, and to test the above theoretical ideas, neutron diffraction measurements have been performed in the three directions of the applied field. 

The refinements obtained from the data collection give magnetic structures which derive from the zero field magnetic structures $\psi_2$ and $\psi_3$ (see Figure \ref{Figure1}) of the $\Gamma_5$ representation. Depending on the strength and on the field direction, the magnetic moments are tilted with respect to the zero field structures, resulting in a non zero magnetic component along the applied field (see Appendix~\ref{app_H}). They still can be related to the $\psi_2$, respectively $\psi_3$, state provided the 4 moments of a tetrahedron have the same component along the local ${\boldsymbol a_i}$, respectively ${\boldsymbol b_i}$, axis. 

When the field is low enough, the field induced tilted $\psi_2$-like state becomes hardly distinguishable from the $\psi_2$ structure so that the discussion in terms of the $n=0,..,5$ domains defined in Section \ref{theory} remains meaningful. Two different phenomena can take place, the {\it domain selection} and/or a {\it phase transition} corresponding to a change of the magnetic structure.

It is worth noting that careful neutron diffraction measurements are necessary to disentangle these two phenomena, requiring one to collect a large number of Bragg peaks integrated intensities. Of course, the refinements also have to take into account the presence of several domains, since the magnetic intensities of the Bragg peaks can be extremely different depending on the domain which is considered, as reported in Table~\ref{table-dom} for the six domains of the $\psi_2$ state and a selection of Bragg peaks.

\begin{table}
\setlength{\extrarowheight}{1pt}
\begin{tabularx}{8.5cm}{*{5}{Y}}
\hline
\hline
$Q$ & $n=0,3$ & $n=1,4$ & $n=2,5$  & 6 domains average\\
\hline 
$(002)$	& 0	& 0	& 0 & 0 \\
$(111)$	& 4	& 4	& 4 & 4 \\
\hline 
$(0,2,2)$ & 2.66	& 10.66	& 2.66	& 5.33 \\
$(2,0,2)$ & 2.66 & 2.66  & 10.66 & 5.33 \\
$(2,2,0)$ & 10.66	& 2.66 & 2.66	  & 5.33 \\
\hline 
$(0,2,\bar{2})$ & 2.66 & 10.66 & 2.66 & 5.33 \\
$(2,0,\bar{2})$ & 2.66 & 2.66 & 10.66 & 5.33 \\
$(2,\bar{2},0)$ & 10.66 & 2.66 & 2.66 & 5.33 \\
\hline 
$(\bar{3},1,1)$ & 3.03 & 0.12 & 3.03 & 2.06 \\
$(1,\bar{3},1)$ & 3.03 & 3.03 & 0.12 & 2.06 \\
$(1,1,\bar{3})$ & 0.12 & 3.03 & 3.03 & 2.06 \\
\hline 
$(\bar{3},\bar{1},1)$& 3.03 & 0.12 & 3.03 & 2.06 \\
$(1,\bar{3},\bar{1})$ & 3.03 & 3.03 & 0.12& 2.06 \\
$(\bar{1},1,\bar{3})$ & 0.12 & 3.03 & 3.03 & 2.06 \\
\hline
$(3 \bar{3} 1)$   & 1.75 & 3.44 & 3.44 & 2.88 \\
\hline \hline
\end{tabularx}
\caption{Magnetic intensities (in arbitrary units) of several Bragg peaks of the different $\psi_2$ domains. $n$ is the integer defining the angle $\phi=n \pi /3$. Note that neutron diffraction cannot distinguish $180^{\circ}$ domains.}
\label{table-dom}
\end{table}

To determine precisely the value of the characteristic fields, we have used magnetization measurements, which allow a better field resolution, and for which, in our conditions, the demagnetization effects are negligible. Nevertheless, the magnetization shows only weak anomalies, and it is necessary to use its derivative. A characteristic field is then defined as the maximum of $dM/dH$ vs $H$ and corresponds to the inflection point in the Bragg peak field dependence. Systematic magnetization measurements have been carried out as a function of temperature to determine the temperature dependence of these characteristic fields. 

In the following, we show the follow-up of some magnetic Bragg peak intensities as a function of magnetic field, obtained for each field direction, as well as sketches of the refined magnetic structures for some values of the magnetic field. While the data are presented in the whole field range, to give a complete picture of the magnetic field effect, the analysis is focused on the low field part of the data, which is of specific interest with respect to the above theoretical predictions.

\begin{figure*}[t]
\includegraphics[width=\textwidth]{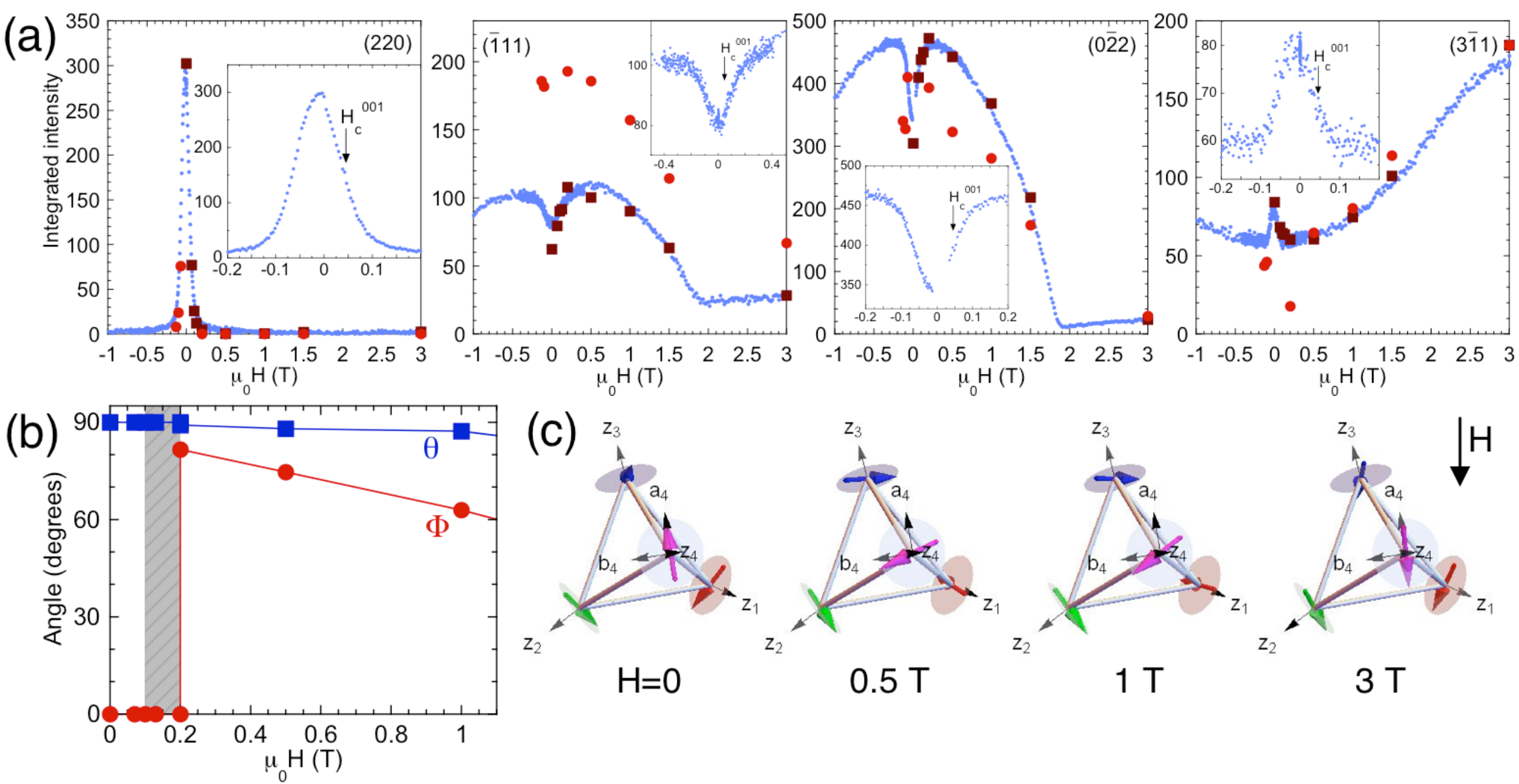}
\caption{(Color online) (a) Field dependence of selected Bragg peaks intensities when sweeping the field from $+3$ to $-3$ T for ${\boldsymbol H} \parallel [001]$ (blue points). The intensity has been rescaled to the integrated intensity obtained in full data collections. The brown squares correspond to data collections and the red dots to the results of the {\sc fullprof} refinements. The insets focus on the low field behavior to emphasize the intensity evolution around $H_c^{001}$. (b) Field dependence of the polar angle describing the spin orientation in its local frame. $\phi$ is the longitude within the XY anisotropy plane and $\theta$ the polar angle. The transition from $\psi_2$ to $\psi_3$ occurs below 0.15 T: for the shown domain, it corresponds to a transition from $\phi=0$ to $\pi/2$. The grey area represents the range in which the refinements are not able to distinguish the $\psi_2$ and $\psi_3$ states. (c) Sketch of the fitted magnetic configurations within a tetrahedron as a function of field. 
}
\label{Figure-001}
\end{figure*}

\begin{figure}[!]
\includegraphics[width=7.5cm]{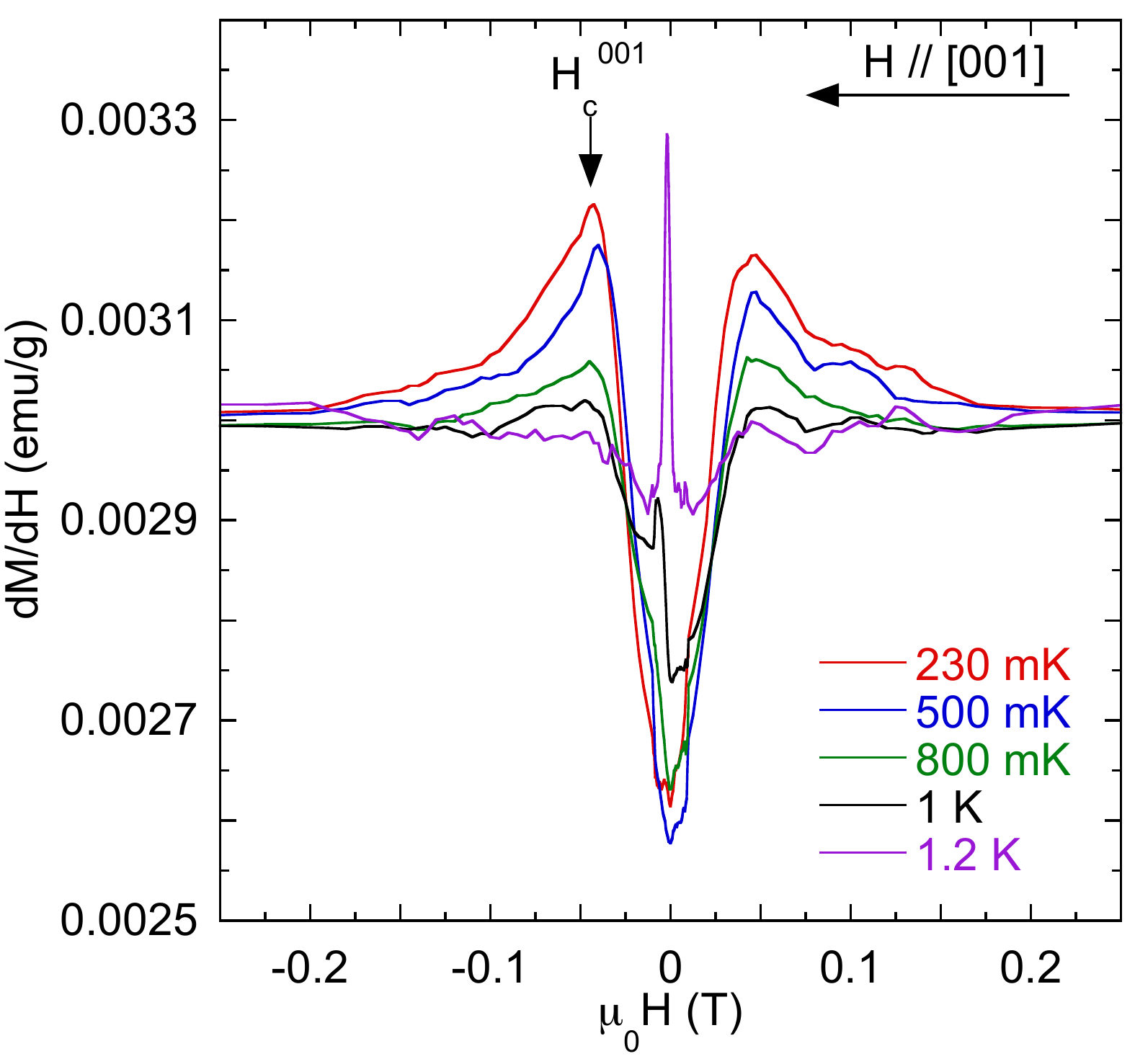}
\caption{(Color online) $dM/dH$ vs $H$ when the field is applied along $[001]$ for several temperatures. The field is swept from positive to negative fields.}
\label{Hc_001}
\end{figure}

\begin{figure*}[t!]
\includegraphics[width=\textwidth]{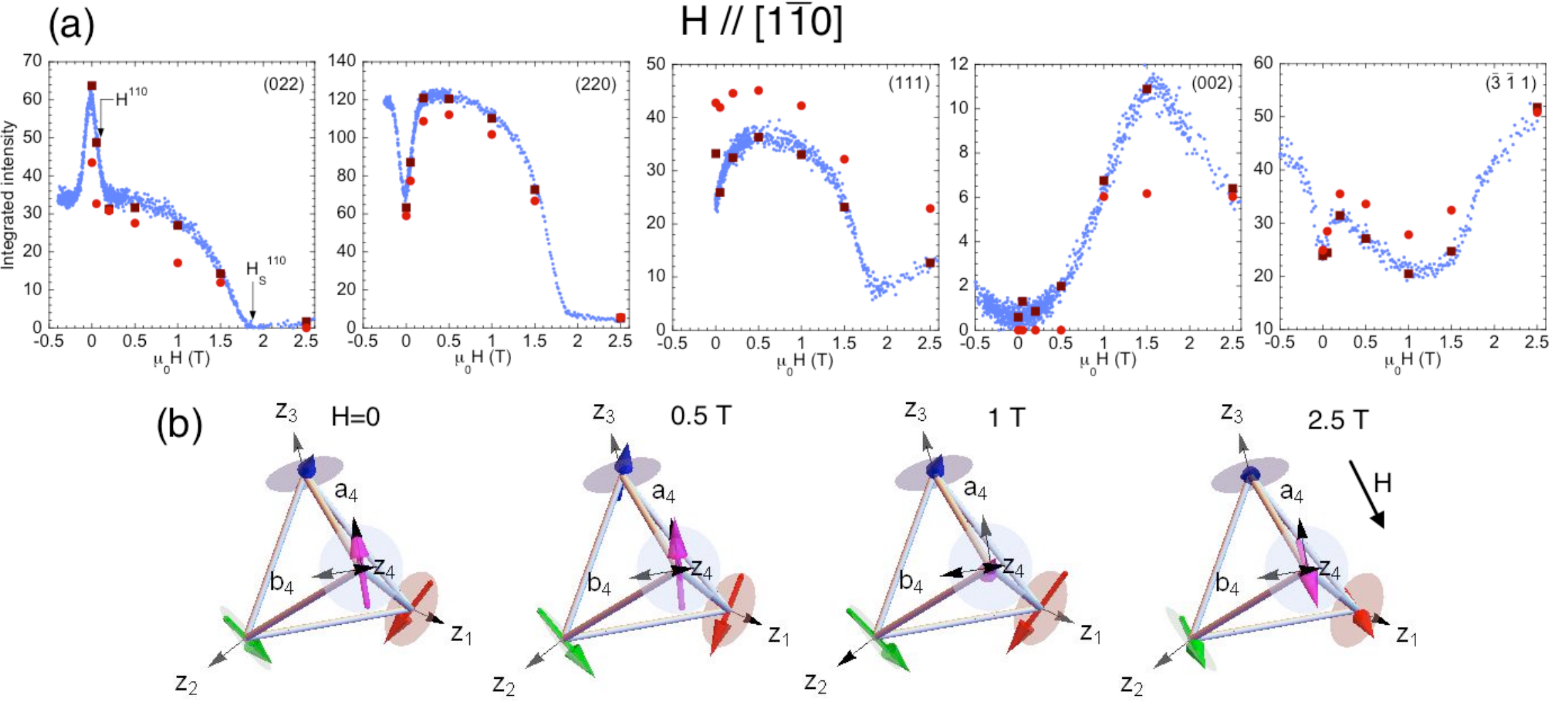}
\caption{(Color online) (a) Field dependence of selected Bragg peaks intensities measured at 60 mK (blue points). The field is varied from 3 down to $-3$ T along the $[1\bar10]$ direction. The intensity has been rescaled to the integrated intensity obtained in full data collections. The brown squares correspond to data collections and the red dots to the results of the {\sc fullprof} refinements. (b) Sketch of the fitted magnetic configurations within a tetrahedron as a function of field. }
\label{Figure-110}
\end{figure*}

\subsubsection{${\boldsymbol H} \parallel [001]$}

Neutron diffraction data indeed point out the predicted phase transition at low field, far from the field polarized state. To evidence this behavior, the field dependence of a few characteristic Bragg peaks, measured at very low temperature ($T= 60$~mK $\ll T_{\rm N}$) is shown in Figure~\ref{Figure-001}(a). Starting from high field (3 T), the $(\bar{1}11)$ and $(0\bar{2}2)$ peaks first show a slope discontinuity at $H_S^{001}$ marking the boundary of the field polarized state. More interestingly, below about 0.15 T, we observe a sharp change of the peak intensity. This effect is even more spectacular for the $(220)$ peak which does not show any field dependence at high field. It is emphasized in the insets of Figure \ref{Figure-001}(a). We can note that the $(\bar{1}11)$ peak varies more smoothly at low field, but its intensity is not sensitive to in-plane reorganizations of the moments.

Our refinements of the data sets at constant field show that these observations correspond to a transition from the $\psi_2$ configuration (see Figure \ref{Figure1}(a)) to a $\psi_3$ state, as predicted (see Appendix \ref{app-001}). Note that in the $\psi_2$ state, the refinements did not allow us to determine the domain population accurately, so that an equipopulation was assumed. Some discrepancies are observed between the measured integrated intensity and the refinement (see for example the $(\bar{1}11)$ Bragg peak in Figure \ref{Figure-001}), mainly on the peaks which have a strong nuclear component. This occurs because the refinements are performed on more than twenty magnetic Bragg peaks. Nevertheless, the shape of the field variation is recovered.  

The transition is further illustrated in Figure \ref{Figure-001}(b) which displays the spherical angles of the magnetic moment at the site labeled 1 in Table \ref{table-sym}. For the sake of clarity, the $n=0$ domain only is represented \cite{footnote}. At low field, $\phi=0$ and $\theta=\pi/2$ (see Figures \ref{Figure1}(a) and \ref{Figure-001}(c) for $H=0$). Above the transition, in contrast, the moments are essentially along the ${\boldsymbol b_i}$ direction, hence $\phi=\pi/2$ (see Figure \ref{Figure-001}(c) for $\mu_0H=0.5$ T). It is found that with increasing field, $\phi$ decreases continuously from $\phi=\pi/2$ down to zero at $H_S^{001}$. $\theta$ also decreases indicating the rise of the weak out-of-plane component. 

This $\psi_2 \rightarrow \psi_3$ transition is similar to a spin-flop mechanism. Starting from the $\psi_3$ configuration, the spins can accommodate the increase of the field along $[001]$ while essentially rotating within their anisotropy plane, hence at low energy cost. This results in an increase of the net moment along the field direction. 
 
The characteristic field $H_c^{001}$ is defined as the maximum of $dM/dH$ vs $H$. This procedure gives the value $\mu_0H_c^{001}=43 \pm 5$ mT at 200 mK. When the temperature is increased, the position of $H_c^{001}$ remains constant within the experimental accuracy. However, above 500~mK, the $dM/dH$ peak broadens and its amplitude continuously decreases until it disappears at the N\'eel temperature $T_N=1.2$ K, as shown in Figure \ref{Hc_001}. This temperature dependence is in strong contrast with the $H_S$ dependence shown in Figure~\ref{pdiag}, and suggests that the ratio between the $A_2$ and $A_6$ coefficients does not change with temperature. Note that a sharp peak appears at zero field at $T_N$. This feature may be related with an anomalous behavior at the transition, but it will be discussed in a forthcoming publication.

\begin{figure}[h!]
\includegraphics[width=7.5cm]{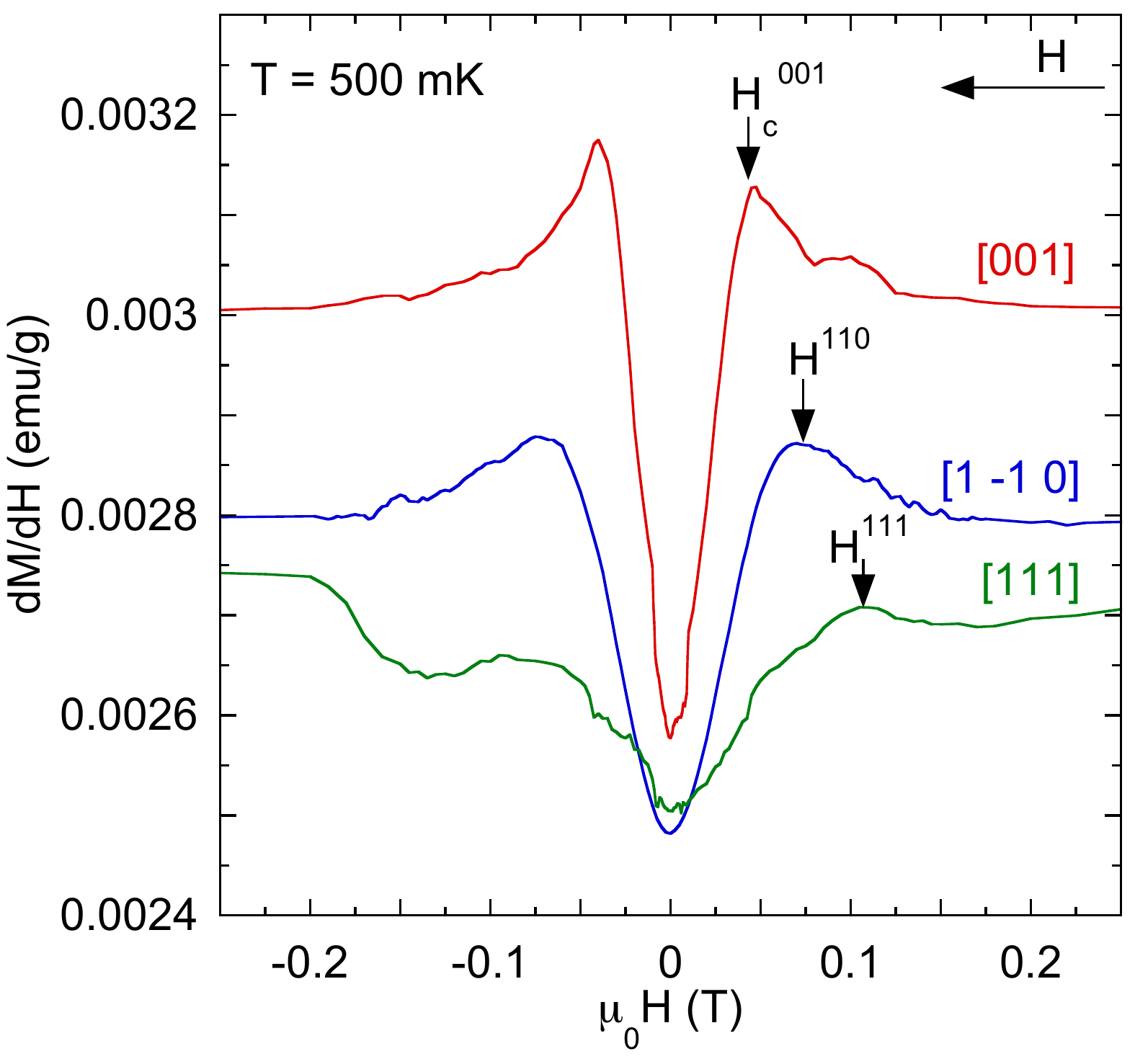}
\caption{(Color online) $dM/dH$ vs $H$ when the field is applied along $[001]$ (red), $[1\bar{1}0]$ (blue) and $[111]$ (green) at 500 mK. The field is swept from positive to negative fields. }
\label{dMdH_3dir}
\end{figure}

\begin{figure*}[t!]
\includegraphics[width=\textwidth]{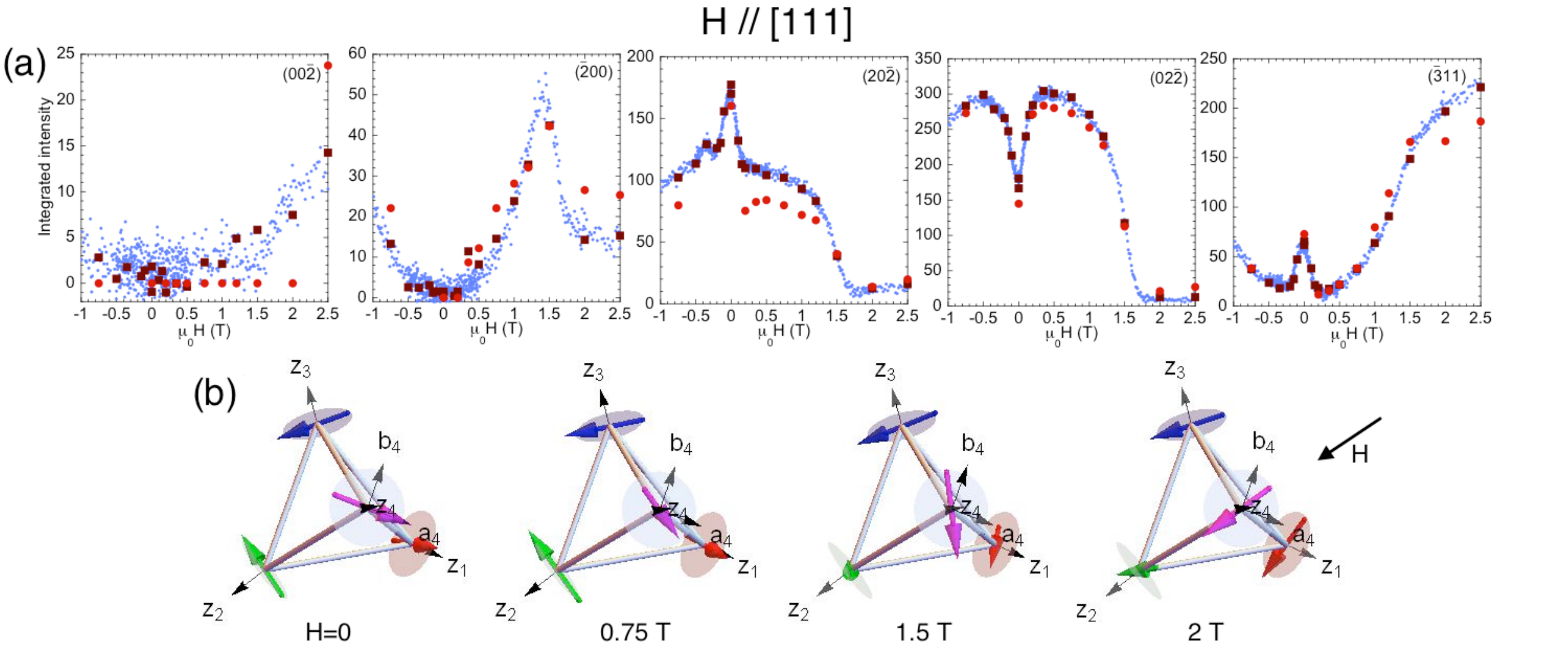}
\caption{(Color online) (a) Field dependence of selected Bragg peaks intensities measured at 60 mK (blue points). The field is varied from 2.5 down to $-2.5$ T along the $[111]$ direction. The intensity has been rescaled to the integrated intensity obtained in full data collections. The brown squares correspond to data collections and the red dots to the results of the {\sc fullprof} refinements. (b) Sketch of the fitted magnetic configurations within a tetrahedron as a function of field. 
}
\label{Figure-111}
\end{figure*}

\subsubsection{${\boldsymbol H} \parallel [1\bar{1}0]$}

Figure \ref{Figure-110}(a) shows the field dependence of selected magnetic Bragg peak intensities measured at $T= 60$ mK. Decreasing field, besides the cusp at $H_S^{1\bar{1}0}$, the data show a smooth evolution, which corresponds to the gradual recovery of the $\psi_2$ structure as previously reported \cite{Ruff08,Cao10}. The slow field sweeping rate allows to highlight a new phenomenon. Below about 0.15 T, strong intensity variations can be seen, upwards or downwards, depending on the Bragg peak. This looks similar to the transition reported above in the $[001]$ direction. It is also observed in the magnetization curves, although less sharply, as shown in Figure \ref{dMdH_3dir} and in Ref. \onlinecite{Petrenko13}. 

Nevertheless, the physical origin is different since, in this direction, no phase transition occurs at low field. The refinements show that the magnetic structure tilts continuously from the $\psi_2$ state up to the transition to the field polarized state at $H_S^{1\bar{1}0}$ (see Appendix \ref{app-1m10}). Actually, this anomalous behavior can be attributed to the selection at finite field of the domains with $\phi=0, \pi$ (i.e. $n=0,3$) among the six possible domains of the $\psi_2$ structure, in agreement with the theoretical analysis presented in section \ref{theory}. Comparing the results of Figure \ref{Figure-110}(a) with the calculations reported in Table~\ref{table-dom}, we find indeed that above about 0.15 T, the intensities are reproduced by the response of the $n=0,3$ domains, while below, they are compatible with the average of 6 equipopulated domains. 

It is also worth noting that some Bragg peaks, like $(\bar{3}\bar{1}1)$ behave differently for positive and negative fields. This may be due to some domain viscosity that we do not understand at the moment. 

Our magnetization data allow us to estimate the characteristic field of this domain selection to $\mu_0 H^{110}=74 \pm 6$~mT (see Figure~\ref{dMdH_3dir}). The temperature dependence of the $dM/dH$ curves is similar to the $[001]$ case, i.e. $H^{110}$ remains constant while the maximum of the derivative gradually disappears when the temperature increases.

\subsubsection{${\boldsymbol H} \parallel [111]$}

To study the $[111]$ case, we follow the same approach and present the field dependence of selected magnetic Bragg peaks intensities measured at $T= 60$~mK (see Figure \ref{Figure-111}). The same qualitative observations are made: when the field decreases from the saturation field, a first regime is observed down to $H_S^{111}$ where a discontinuity is observed on several peaks. Then, the peak intensities show a smooth evolution down to about 0.2~T, where the system enters another regime with a spectacular intensity increase or decrease, which is Bragg dependent. 

Refinements from the data collections performed at constant fields show that, like in the $[1\bar{1}0]$ case, the system remains in a tilted $\psi_2$-like state in the whole field range below $H_S^{111}$, suggesting that the intermediate phase predicted to occur at large $\zeta$ is not stabilized. 

More precisely, the refinements show that, below $H_S$, the magnetic structure is derived from the $\phi=\pi/3$ domain of the $\psi_2$ state only. Below 0.2~T, comparing the results of Figure \ref{Figure-111}(a) with the calculations reported in Table~\ref{table-dom}, we find that the population of the $\phi= \pi$ and $5\pi/3$ domains increases while the $\phi=\pi/3$ domain population decreases. The low field anomalous behavior is thus not due to a phase transition but to a domain selection. 

This observation differs from the theoretical prediction where an equipopulation of the three domains is expected, due to the absence of the three-fold symmetry breaking when the field is exactly applied along $[111]$. Actually, due to a small misalignment of the field (the angles between $[11\bar{2}]$ and $[2\bar{2}0]$ with the horizontal scattering plane were measured to be $-0.22^{\circ}$ and $3.3^{\circ}$ respectively), the $A_2$ term given by Eq. \ref{df2} is reintroduced in the free energy. Simple calculation shows that the domain $\phi=\pi/3$ is favored with respect to the others, in agreement with the experiment. 

\begin{figure}[!]
\includegraphics[width=7.5cm]{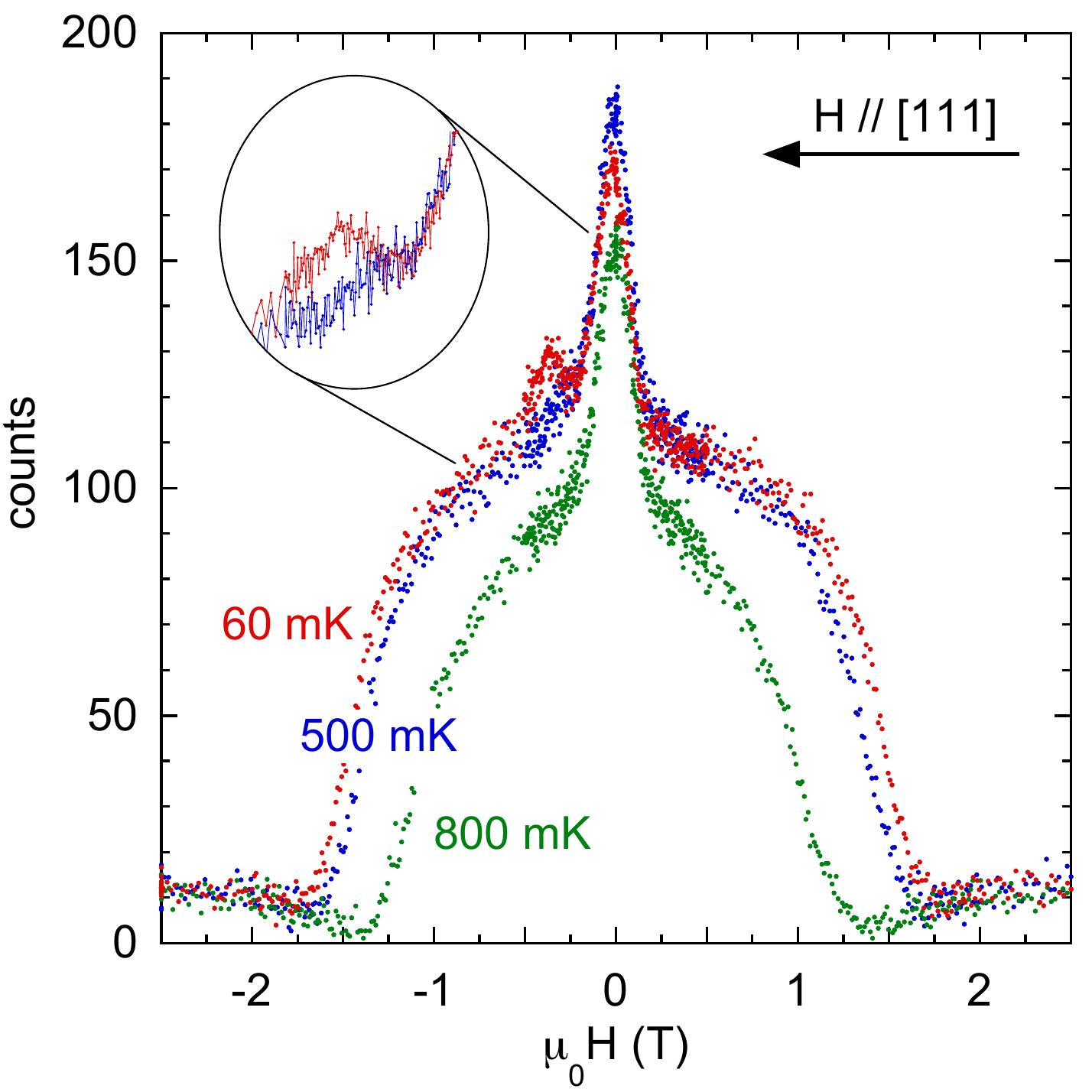}
\caption{(Color online) Magnetic intensity of the $(20\bar{2})$ Bragg peak for ${\boldsymbol H} \parallel[111]$, swept from 2.5 to $-2.5$ T, at several temperatures: 60, 500 and 800~mK. The inset emphasizes the small hysteretic anomaly which disappears at 500 mK. }
\label{111_20m2}
\end{figure}

The odd behavior of the $(20\bar{2})$ Bragg reflection, which exhibits a small bump around 0.35 T, has to be mentioned here. This feature is hysteretic and appears for $H< 0$ (resp. $H >0$) when sweeping the field from the positive (resp. negative) side. Its origin remains unclear, as it is not observed on other Bragg peaks, such as the $(02\bar{2})$ and $(2\bar{2}0)$ ones. When increasing the temperature at 500~mK and above, this hysteretic behavior vanishes as shown in Figure \ref{111_20m2}. 

The domain selection along this $[111]$ direction is hardly measurable in the magnetization, as shown in Figure \ref{dMdH_3dir}. A very large bump is observed in $dM/dH$ at about $0.1 \pm 0.01$ T, which does not move but progressively disappears when the temperature increases. This is consistent with neutron diffraction measurements at 500 and 800~mK, which show that the change of behavior at low field of the magnetic peaks remains at the same characteristic field but becomes smoother when the temperature increases (see Figure \ref{111_20m2}). 

\section{Discussion}
This experimental and systematic low field study is in qualitative agreement with the analysis performed by Maryasin et al. \cite{Maryasin16}. 

When the field is applied along the $[001]$ direction, the magnetic structure determination from neutron scattering measurements confirms the existence of a spin-flop like transition from the $\psi_2$ to the $\psi_3$ state. 

Along the $[1\bar{1}0]$ direction, our analysis shows that the previously observed but unexplained magnetization anomaly is due to an abrupt domain selection and not to a phase transition.

Finally, while it had been proposed as a possible scenario, no phase transition is observed along $[111]$. However, due to a small disorientation of the magnetic field, we cannot conclude definitely on this point. 

Importantly, the characteristic fields do not depend on the temperature up to $T_{\rm N}$. This means that the anisotropy terms are independent of temperature, or have all the same temperature dependence, which is not what is expected a priori, since these terms depend on different powers of the magnetic order parameter. 

\begin{table}
\setlength{\extrarowheight}{1pt}
\begin{tabularx}{8.5cm}{*{3}{Y}}
\hline
\hline
 & Exp & Calc \\
\hline
$\Delta$ & 43 $\mu$eV & 15-20~$\mu$eV \\
\hline
$A_6$ & $0.605~\mu$eV & $0.068~\mu$eV \\
$A_2$ & 0.185~meV/T$^2$ & 0.021 meV/T$^2$\\
$A'_6$ & - & 0.0027 meV/T$^2$\\
\hline
\hline
\end{tabularx}
\caption{Experimental and theoretical anisotropy parameters. The following microscopic parameters have been used ${\sf J}_{\pm} \approx 0.06 \pm 0.005$ meV, ${\sf J}_{\pm\pm} \approx 0.043 \pm 0.002$ meV and $g_{\perp} \approx 6$ \cite{Savary12,Zhitomirsky12,Petit14}. The ``experimental values'' for $A_6$ and $A_2$ are calculated based on the equations given in the main text and assuming that ${\sf J}_{\pm}$ and ${\sf J}_{\pm\pm}$ are correct.}
\label{compar}
\end{table}

From the characteristic magnetic fields determined in this study, it is now possible to make a quantitative comparison with the theoretical estimations. First, theoretically, $H^{110}/H_c^{001}=\sqrt{2}$. Experimentally we get $H^{110}/H_c^{001}\sim 1.7$, which is slightly larger although in the same range. It is possible to go further by considering the expression of the anisotropy terms in the context of the microscopic Hamiltonian proposed for \erti\ \cite{Maryasin14}. The 6-fold anisotropy of $\Delta F_o$ is given by:
$$A_6 = A^{\rm Th}_6 + A^{\rm ObD}_6$$
with
\begin{eqnarray*}
A^{\rm Th}_6 &=& \frac{N}{216} \frac{k_B T {\sf J}^3_{\pm\pm}}{{\sf J}^3_{\pm}} = \frac{N}{27}~\epsilon^3 ~k_B T \\
A^{\rm ObD}_6 &\sim & \frac{NS}{192} \frac{ {\sf J}^3_{\pm\pm}}{{\sf J}^2_{\pm}} = \frac{N}{48}~\epsilon^3~{\sf J}_{\pm}
\end{eqnarray*}
and the anisotropies relevant in the presence of an applied field write \cite{Maryasin16}:
\begin{equation*}
A_2 = \frac{N}{96} \frac{(g_{\perp} \mu_B)^2}{{\sf J}_{\pm}}~\mbox{and}~ 
A'_6 = \epsilon^2 A_2 ~~\mbox{with}~\epsilon = \frac{1}{2} \frac{{\sf J}_{\pm \pm}}{{\sf J}_{\pm}}
\end{equation*}
With the following parameters ${\sf J}_{\pm} \approx 0.06 \pm 0.005$ meV, ${\sf J}_{\pm\pm} \approx 0.043 \pm 0.002$ meV and $g_{\perp} \approx 6$, we are now in position to compare theory and experiment (see Table~\ref{compar}). 
The calculated $H_c^{001}$ gives a value of about 170mT. Following Ref. \onlinecite{Maryasin16}, this value would correspond to the foot of the (220) peak in Figure \ref{Figure-001}(a), which is hard to determine accurately from an experimental point of view. We can estimate it at about 100~mT, while the characteristic field from the inflection point criteria is 43~mT. The calculated $H_c^{001}$ (and so $H^{110}$) thus overestimates the measured value.
Since the measurements of the characteristic fields only give access to the ratio $A_6/A_2$, it is not possible to determine whether $A_6$, $A_2$ or both is responsible for this discrepancy. Nevertheless, this result would tend to indicate that the $A_2$ term is underestimated theoretically. Nevertheless, the $A'_6$ parameter, which has not been considered in the present approach, may also affect this result.

We can now compare the neutron diffraction and magnetization measurements results with the inelastic neutron scattering experiments. Indeed, as previously discussed, the $A_6$ anisotropy term is responsible for the $\psi_2$ magnetic ordering in zero field, and gives rise to a gap $\Delta$ in the spin wave excitations. From the above study, it is expected that the other terms will play a role in the presence of a magnetic field. It is thus of interest to focus on the spin dynamics in this low field regime, to get a better insight in these terms.
Using the theoretical background developed in Ref.~\onlinecite{Savary12}, the evolution of the spin gap $\Delta$ at low field can be determined. For a field applied along $[1\bar{1}0]$, it writes (see Appendix \ref{models}):
\begin{equation}
\Delta \approx \sqrt{54 A_6 {\sf J}_{\pm} + 3 A_2 H^2 {\sf J}_{\pm}}\\
\label{gaph}
\end{equation}
Fitting our measurements (see Figure \ref{Figure-INS}(b)), we obtain: $54 A_6 {\sf J}_{\pm}=1.96 \pm 0.02 \times 10^{-3}$ meV$^{2}$ and $ 3 A_2 {\sf J}_{\pm}=0.033 \pm 0.003$ meV$^{2}/$T$^{2}$ which leads to: 
$$A_6=0.605 \times 10^{-3} \textrm{ meV \quad and \quad } A_2=0.185 \textrm{ meV/T}^2$$
These coefficients are larger than the estimation from the microscopic Hamiltonian in the order by disorder scenario (see Table \ref{compar}), consistently with our above conclusion that $A_2$ is underestimated by the theory. The $A_6/A_2$ ratio, however, leads to $\mu_0H_c^{001}=170$ mT, in agreement with the theoretical prediction but does not match with the diffraction and magnetization data. This suggests that the expressions used for $\mu_0H_c^{001}$ and $\mu_0H^{110}$ (see Table \ref{summary}) to describe the field induced structures are too simplified and may omit some terms which are not negligible. This discrepancy may also suggest that some ingredients are missing in the Hamiltonian to describe accurately \erti\, and in this sense, fits the conclusions drawn from the field dependence of the spin gap below $H_S$.

\section{Conclusion}

By combining neutron diffraction and magnetization measurements, we have shown that a field induced transition occurs when the field is applied along the $[001]$ direction, as theoretically predicted. In the other directions, only domain selection occurs, which manifests as a sharp variation of the magnetic intensities for some Bragg peaks. However, the characteristic fields we have observed are lower than the theoretical predictions, suggesting that the field induced anisotropy terms are stronger than predicted. 

The characteristic fields do not depend on temperature: when increasing the temperature, the anomalies broaden but remain at the same position before being suppressed at the N\'eel temperature. This is a puzzling result which indicates that the key parameter of the transition and domain selections, the $A_6/A_2$ ratio, remains constant in the whole ordered regime. 
Further theoretical studies are thus needed to quantitatively understand our observations.

Finally, the field dependence of the spin gap is not reproduced by the LSW and VCF models, pointing out that additional terms, such as multispin interactions, may have to be considered in the Hamiltonian to describe the field induced properties.

\medskip
{\it Note added.} 
We were recently aware of a publication appeared in Phys. Rev. B {\bf 95}, 054407 (2017) about the same topic. We basically agree on the existence of a transition for a field along $[001]$. However, this study reports a transition field of 0.18 T, thus  a larger field than our study. It is worth noting that both works differ from the methodology: we analyze and fit diffraction data while this work reports on the evolution of the elastic reponse at the 220 Bragg position measured by time-of-flight neutron scattering at several magnetic fields. In a $[111]$ field, the same study reports on a series of ``domain-based phase transitions'' for fields of 0.15 and 0.40 T whereas we find only domain selection effects below 0.1 T.


\acknowledgments{We thank C. Paulsen for allowing us to use his SQUID dilution magnetometers. We also would like to acknowledge M. Gingras, M. Zhitomirsky and R. Ballou for fruitful discussions. Finally, we acknowledge J. Debray for his help in the orientation of the single crystals.}

\appendix

\section{Magnetic structures}
\label{app_H}
\subsection{Magnetic structure for ${\boldsymbol H} \parallel [001]$}
\label{app-001}

Below $H_c^{001}$, the refinements were carried out using the 6 domains of the $\psi_2$ configuration belonging to the $\Gamma_5$ irreducible representation. Above $H_c^{001}$, the structure was refined in the following model: 
\begin{eqnarray*}
{\boldsymbol m_1} &=& m_x {\boldsymbol a_1}(n=0) + m_y {\boldsymbol b_1}(n=0) + m_z {\boldsymbol z_1} \\
{\boldsymbol m_2} &=& m_x {\boldsymbol a_2}(n=0) + m_y {\boldsymbol b_2}(n=0) + m_z {\boldsymbol z_2} \\
{\boldsymbol m_3} &=& -m_x {\boldsymbol a_3}(n=0) + m_y {\boldsymbol b_3}(n=0) - m_z {\boldsymbol z_3} \\
{\boldsymbol m_4} &=& -m_x {\boldsymbol a_4}(n=0) + m_y {\boldsymbol b_4}(n=0) - m_z {\boldsymbol z_4}
\end{eqnarray*}
where $(m_x,m_y,m_z)$ are the fitted parameters. At $H_c^{001}$, $m_x=0$. With increasing the field, however, $m_y$ weakens while $m_x$ and $m_z$ increases. $m_y$ becomes exactly zero above $H_S^{001}$. Note that the two 180$^{\circ}$ domains cannot be distinguished by neutron scattering and that the net moment within one tetrahedron is along ${\boldsymbol H}$.

\subsection{Magnetic structure for ${\boldsymbol H} \parallel [1\bar{1}0]$}
\label{app-1m10}

For ${\boldsymbol H} \parallel [1\bar{1}0]$ and below $H^{110}$, the refinements were conducted using the 6 $\psi_2$ domains. Above $H^{110}$, the structure was refined in the following model: 
\begin{eqnarray*}
{\boldsymbol m_1} &=& m_x {\boldsymbol a_1}(n=0) - m_y {\boldsymbol b_1}(n=0) \\
{\boldsymbol m_2} &=& m_x {\boldsymbol a_2}(n=0) + m_y {\boldsymbol b_2}(n=0) \\
{\boldsymbol m_3} &=& m'_x {\boldsymbol a_3}(n=0) - m_z {\boldsymbol z_3} \\
{\boldsymbol m_4} &=& m''_x {\boldsymbol a_4}(n=0) + m'_z {\boldsymbol z_4}
\end{eqnarray*}
where $(m_x,m'_x,m''_x,m_y,m_z,m'_z)$ are the fitted parameters. Below 0.5~T, we assumed $m'_z=m_z$ and $m''_x=m'_x$, while above these values could be different. 
Above $H_S^{110}$, $m_x=0$, $m''_x=m'_x$ and $m'_z=m_z$. This model is 
$\psi_2$-like provided that $m_x, m'_x$ and $m''_x$ have the same sign.

\subsection{Magnetic structure for ${\boldsymbol H} \parallel [111]$}
\label{app-111}

For ${\boldsymbol H} \parallel [111]$ the structure was refined in the following model: 
\begin{eqnarray*}
{\boldsymbol m_1} &=& m_{x1} {\boldsymbol a_1}(n=2) + m_{y1} {\boldsymbol b_1}(n=2) \\
{\boldsymbol m_2} &=& m_{x2} {\boldsymbol a_2}(n=2) + m_{y2} {\boldsymbol b_2}(n=2) + m_z {\boldsymbol z_2} \\
{\boldsymbol m_3} &=& m_{x3} {\boldsymbol a_3}(n=2) + m_{y3} {\boldsymbol b_3}(n=2) \\
{\boldsymbol m_4} &=& m_{x4}{\boldsymbol a_4}(n=2) + m_{y4} {\boldsymbol b_4}(n=2) 
\end{eqnarray*}
where $(m_{xi},m_{yi}, m_z)$ are the fitted parameters. This model is 
$\psi_2$-like provided $m_{xi}$ have the same sign which is found to be the case below $H_S^{111}$. At low field, below 0.2~T, the refinements were carried out in the standard $\psi_2$ configuration.

\section{Models}
\label{models}
In this appendix, we describe the two models used to calculate the field dependence of the spin gap $\Delta$. 

Following Ref. \onlinecite{Savary12}, \erti\, is described by a bilinear quadratic Hamiltonian ${\cal H}$ written in terms of the components of an effective spin 1/2 spanning the subspace of the ground \er\, crystal field doublet:
\begin{eqnarray*}
{\cal H} &=& \sum_i g~{\sf S}_i.{\boldsymbol H} + \frac{1}{2} \sum_{i,j} {\sf J}_{zz} {\sf S}^z_i {\sf S}^z_j + {\sf J}_{z \pm}{\sf S}_i^z \left( \zeta_{ij} {\sf S}^+_j + \zeta^*_{ij} {\sf S}^-_j\right) \\
& &
+ 
{\sf J}_{\pm\pm} \left(\gamma_{ij} {\sf S}^+_i {\sf S}^+_j + \gamma^*_{ij} {\sf S}^-_i {\sf S}^-_j \right) - {\sf J}_{\pm} \left({\sf S}^+_i {\sf S}^-_j + {\sf S}^-_i {\sf S}^+_j \right)
\end{eqnarray*}
${\sf S}_i$ denote the pseudo spin 1/2 written in its local basis spanned by the site dependent $({\boldsymbol a_i},{\boldsymbol b_i},{\boldsymbol z_i})$ (see Table \ref{table-sym}). $g$ is an effective anisotropic tensor and $({\sf J}_{\pm\pm},{\sf J}_{\pm},{\sf J}_{z\pm},{\sf J}_{zz})$ is a set of effective exchange parameters allowed by symmetry. Fitting the spin wave excitations \cite{Savary12} leads to:
\begin{eqnarray*}
{\sf J}_{zz} &=& -2.5 \pm 1.8~\times~10^{-2}~\textrm{meV}\\
{\sf J}_{z \pm} &=& -0.88 \pm 1.5 ~\times~10^{-2}~\textrm{meV}\\
{\sf J}_{\pm\pm} &=& 4.2 \pm 0.5 ~\times~10^{-2}~\textrm{meV}\\
{\sf J}_{\pm}&=& 6.5 \pm 0.75 ~\times~10^{-2}~\textrm{meV}
\end{eqnarray*}
in combination with:
$$ g = \left(
\begin{array}{ccc}
5.97\pm 0.08 & & \\
& 5.97\pm 0.08 & \\
& & 2.45 \pm 0.23
\end{array}
\right)$$
The spin wave spectrum is calculated following Ref. \onlinecite{spinwave} using a Bogoliubov transform. In this approach, the spin gap can be calculated analytically introducing an effective anisotropy term $-\frac{\lambda}{2} \cos{6\phi}$. Expanding the cosine to second order in $\phi$, the authors of Ref. \onlinecite{Savary12} obtain:
\begin{equation*}
\Delta = \sqrt{18 \frac{\lambda}{\eta}}
\end{equation*}
with $\eta=\frac{4}{3}\frac{1}{2{\sf J}_{\pm}+{\sf J}_{z z}}$, hence
\begin{equation*}
\Delta \approx \sqrt{27 \lambda {\sf J}_{\pm}}
\end{equation*}
In Ref. \onlinecite{Savary12}, this $\cos{6\phi}$ dependence is obtained numerically, by computing the contribution to the total energy of the spin wave zero point energy. This calculation is done for a number of magnetic structures  described by $\phi$, with $0 \leq \phi \leq 2 \pi$. With the notations of the present work, we write $\frac{\lambda}{2}=A_6$ to obtain:
\begin{equation*}
\Delta \approx \sqrt{54 A_6 {\sf J}_{\pm}}
\end{equation*}
For a magnetic field ${\boldsymbol H} \parallel [1\bar{1}0]$, the anisotropy term writes $- A_6 \cos 6\phi - \frac{1}{2} A_2 H^2 \cos 2\phi$. Expanding this expression to second order in $\phi$, we find that $A_6$ is replaced by $A_6+ \frac{A_2 H^2}{18}$, hence:
\begin{equation*}
\Delta \approx \sqrt{54 A_6 {\sf J}_{\pm} + 3 A_2 H^2 {\sf J}_{\pm}}\\
\end{equation*}

\begin{table}[!t]
\begin{tabular}{*{3}{p{2.5cm}}}
\hline
\hline
Coupling & Ref. \onlinecite{Savary12} & VCF model\\ 
\hline
${\sf J}_{\pm \pm}$ & 4.2 $\pm$ 0.5 & 4.45 $\pm$ 0.1\\
${\sf J}_{\pm}$   & 6.5 $\pm$ 0.75 & 5.85 $\pm$ 0.1\\
${\sf J}_{z\pm}$  & -0.88 $\pm$ 1.5 & 0.92 $\pm$ 0.1\\
${\sf J}_{zz}$    & -2.5 $\pm$ 1.8  & -0.87 $\pm$ 0.1\\
\hline
\hline
\end{tabular}
\caption{Anisotropic exchange parameters. Units are in $10^{-2}$ meV. }
\label{param}
\end{table}

The ``VCF'' model proceeds differently as explained in Ref. \onlinecite{Petit14} and \onlinecite{Rau16}. The Hamiltonian is written in terms of the actual magnetic moments ${\boldsymbol J}_i$ (written in the cubic global frame) and takes into account explicitly the CEF Hamiltonian ${\cal H}_{\rm CEF}$ \cite{Cao09, Bertin12,Cao10,Bonville13}:
\begin{equation}
{\cal H}_{\rm VCF} = {\cal H}_{\rm CEF} + \sum_i g_J \mu_{\rm B} {\boldsymbol J}_i.{\boldsymbol H} + \frac{1}{2} \sum_{i,j} {\boldsymbol J}_i {\cal J}_{i,j} {\boldsymbol J}_j
\end{equation}
The convention here is to define ${\cal J}_{i,j}$ in the $({\boldsymbol a},{\boldsymbol b},{\boldsymbol c})$ frame linked with a R-R bond:
\begin{eqnarray*}
{\boldsymbol J}_i \cdot {\cal J}_{i,j} \cdot {\boldsymbol J}_j &=& \sum_{\mu,\nu=x,y,z} J_i^{\mu} 
\left( 
{\cal J}_a a_{ij}^{\mu} a_{ij}^{\nu} + 
{\cal J}_b b_{ij}^{\mu} b_{ij}^{\nu} \right.\\
& & \left.
+ {\cal J}_c c_{ij}^{\mu} c_{ij}^{\nu} 
\right) J_j^{\nu} + {\cal J}_4 \sqrt{2}~{\boldsymbol b_{ij}}.({\boldsymbol J}_i \times {\boldsymbol J}_j)
\end{eqnarray*}
Considering for instance the pair of \er\, ions at ${\boldsymbol r_1}=(1/4,3/4,0)a$ and ${\boldsymbol r_2}=(0,1/2,0)a$, where $a$ is the cubic lattice constant, we define the local bond frame as: ${\boldsymbol a_{12}} = (0,0,-1)$, ${\boldsymbol b_{12}} = 1/\sqrt{2} (1,-1,0)$ and ${\boldsymbol c_{12}} = 1/\sqrt{2} (-1,-1,0)$. This Hamiltonian, written in terms of bond-exchange constants, has the advantage to provide a direct physical interpretation of the different parameters. Note that ${\cal J}_4$ is an anti-symmetric exchange constant (Dzyaloshinskii-Moriya like), while ${\cal J}_{a,b,c}$ are symmetric terms. Fitting the spin wave excitations in the RPA approximation \cite{jensen} leads to \cite{Petit14}:
\begin{eqnarray}
{\cal J}_a & \sim & 0.003 \pm 0.005~{\rm K} \quad {\cal J}_b \sim 0.075 \pm 0.005 ~{\rm K} \nonumber \\
{\cal J}_c & \sim & 0.034 \pm 0.005~{\rm K} \quad {\cal J}_4 \sim 0 \pm 0.005 ~{\rm K} \nonumber .
\label{Kresults}
\end{eqnarray}
To compare these exchange parameters  with those of Ref. \onlinecite{Savary12}, the VCF Hamiltonian can  be projected onto the spin components of the pseudospin ${\sf S}_i$ using the effective $g$-tensor ${\sf J} = g/g_J{\sf S} = \lambda{\sf S}$, leading to the transformed couplings:
\begin{eqnarray*}
{\sf J}_{zz}   & = & \lambda_z^2 ~\frac{{\cal J}_a-2{\cal J}_c-4{\cal J}_4}{3} \\
{\sf J}_{\pm}  & = & -\lambda_{\perp}^2 ~\frac{2{\cal J}_a-3{\cal J}_b-{\cal J}_c+4{\cal J}_4}{12} 
\end{eqnarray*}
\begin{eqnarray*}
{\sf J}_{z\pm}  & = & \lambda_{\perp}~\lambda_z ~\frac{{\cal J}_a+{\cal J}_c-{\cal J}_4}{3 \sqrt{2}} \\
{\sf J}_{\pm \pm} & = & \lambda_{\perp}^2 ~\frac{2{\cal J}_a+3{\cal J}_b-{\cal J}_c+4{\cal J}_4}{12}
\end{eqnarray*}
With the Wybourne coefficients that enter ${\cal H}_{\rm CEF}$ proposed in Ref. \onlinecite{Cao09}, the effective $g$-tensor writes:$$ g = \left(
\begin{array}{ccc}
6.78 & & \\
& 6.78 & \\
& & 2.73
\end{array}
\right)$$ and the transformed parameters are given in Table \ref{param}.




\begin{thebibliography}{99}
\bibitem{bnm} The crystal field is modeled by the following coefficients: $B_{20} = 616$ K, $B_{40} = 2850$ K, $B_{43} = 795$ K, $B_{60} = 858$ K, $B_{63} = -493$ K, $B_{66} = 980$ K, in Wybourne conventions.
\bibitem{footnote} The $n=0$ domain is in principle not favored by a field along $[001]$ as previously discussed. However, since the experimental data were not able to refine the population of the different domains at low field, we have chosen to present the case of the $n=0$ domain for the sake of simplicity.
\end{thebibliography}
\end{document}